**Intrinsic hole mobility and trapping in a regio-regular poly(thiophene)**

A. Salleo, T. W. Chen, A. R. Völkel

Palo Alto Research Center, 3333 Coyote Hill Road, 94304 Palo Alto, CA

Y. Wu, P. Liu, and B. S. Ong

Materials Design & Integration Laboratory, Xerox Research Centre of Canada Mississauga, Ontario, Canada L5K2L1

and R. A. Street

Palo Alto Research Center, 3333 Coyote Hill Road, 94304 Palo Alto, CA

**Abstract**

The transport properties of high-performance thin-film transistors (TFT) made with a regio-regular poly(thiophene) semiconductor (PQT-12) are reported. The room-temperature field-effect mobility of the devices varied between 0.004 cm$^2$/V s and 0.1 cm$^2$/V s and was controlled through thermal processing of the material, which modified the structural order. The transport properties of TFTs were studied as a function of temperature. The field-effect mobility is thermally activated in all films at T<200 K and the activation energy depends on the charge density in the channel. The experimental data is compared to theoretical models for transport, and we argue that a model based on the existence of a mobility edge and an exponential distribution of traps provides the best interpretation of the data. The differences in room-temperature mobility are attributed to different widths of the shallow localized state distribution at the edge of the valence band due to structural disorder in the film. The free carrier mobility of the mobile states in the ordered regions of the film is the same in all structural modifications and is estimated to be between 1 and 4 cm$^2$/V s.

## 1-Introduction

The carrier mobility of polymer semiconductors has improved tremendously over the past few years. Field-effect mobility as high as 0.1 cm$^2$/V s has recently been measured in regio-regular poly(thiophenes).[1-3] Transport characteristics are strongly dependent on the degree of order of the polymer semiconductor at the dielectric interface. Because the structure of the polymer depends on the processing conditions, it is not uncommon to find in the literature widely differing mobility values obtained for nominally the same polymer. In particular, the energy of the dielectric surface prior to the deposition of the polymer,[4-7] the solvent evaporation rate,[2] the molecular weight of the polymer[8] and thermal post-processing of the film[9] all influence the carrier mobility.

There is no general consensus on the mechanism of charge transport in these amorphous or poly-crystalline materials. A complete model of the electrical properties should include a description of the energy distribution of the carriers and how the conduction varies as a function of energy. Disorder-induced localized states are clearly important for the transport, and the essential question is the relation between atomic structure, electronic structure and the transport. Generally, charge transport in disordered materials is described either as hopping between localized states, or trapping and release from localized states into a higher energy mobile state.[10] The degree of structural order may change the mechanism even within the same class of polymer.

Because the electronic structure of semiconducting polymer films is not known experimentally, a simplified model has to be assumed. The model proposed by Bässler[11] assumes that the energy distribution of the carriers is Gaussian due to the random disorder in the material. The standard deviation of the Gaussian (typically 0.1 eV) increases with increasing disorder in the material. To simplify the calculations, electronic structures comprising an exponential tail in the bandgap are often used as well. Tanase et al.[12] have unified these two descriptions by showing that for field-effect transistors (FETs), where the areal charge density in the semiconductor is high, an exponential density-of-states (DOS) is a good approximation of the Gaussian DOS.

The aim of this paper is to explore the relation between structural order and electronic conduction in polythiophene thin film transistors, in order to understand the electronic structure and transport mechanisms. We modify the polymer structure by different thermal processing, and can vary the mobility by a factor of 25. The carrier mobility also depends on the drift field, the carrier density and temperature.[13-20] In field-effect transistors (FETs) operating in the linear regime, the dependence of the mobility on the drift field can be neglected since the geometry of the device generates small lateral fields (~ few $10^4$ V/cm) for typical device geometries (channel length>5 μm). We use the dependence of the mobility on carrier density and temperature to determine the transport mechanism and other properties of the polymer, such as its free carrier mobility and the shape of the DOS. In polymers however, transport measurements are limited to a narrow temperature range because of material degradation at moderate temperatures, which makes it difficult to test transport theories.

Section 2 describes experimental results, obtained by measuring the characteristics of regio-regular poly(thiophene) transistors as a function of temperature for different processing conditions. Section 3 compares a variable range hopping model with a mobility edge (ME) model with an exponential density of shallow traps. In section 4 we argue that the ME model is preferred, and extract the materials parameters accordingly. The I-V curves of our devices as a function of temperature are simulated using these parameters and compared to the experimental curves. We estimate the carrier mobility of the mobile states, and show that the variation of mobility with different processing conditions can be attributed to a different extent of trapping in disordered states. The free carrier mobility in all the devices is approximately constant and we estimate it to be between 1 and 4 $cm^2$/V s.

**2-Experimental**

2-1 Device fabrication and characterization

Coplanar thin-film transistors (TFTs) were fabricated on a doped Si wafer that acted as a common gate electrode. The gate dielectric was a 100 nm layer of thermally grown oxide coated with a monolayer of octyltrichlorosilane or octadecyltrichlorosilane.[6] The surface treatment of the dielectric improved the mobility of the transistor by approximately three orders of magnitude. The devices were made by spin-coating on a chip with pre-patterned bottom Au contacts a solution of poly[5,5'-bis(3-alkyl-2-thienyl)-2,2'-bithiophene)] (PQT-12), a poly(thiophene) synthesized by Xerox Research Centre of Canada.[21] The monomer of PQT-12 has symmetric alkyl side chains, which ensures that the resulting polymer is regio-regular. The channel lengths varied between 40-100 µm and the widths varied between 500-1000 µm. The semiconductor thickness varied between 20-60 nm.

The TFTs were electrically characterized by measuring transfer and output curves in a vacuum probe station (P<1mTorr) equipped with a Joule-Thomson refrigerating stage for low-temperature measurements down to ~80 K (MMR Technologies). Two measurement modes were used. In the DC mode, the gate electrode was biased continuously during the voltage sweep. In this mode, the measurement of a transfer curve took several minutes. Bias stress due to trapping of carriers at the semiconductor/dielectric interface was occasionally observed at low temperature after taking measurements in the DC mode. Detrapping of carriers at room temperature in PQT-12 is relatively fast (~1 s) but can take longer at lower temperatures.[3] In order to effectively guarantee threshold voltage stability over the course of the temperature ramp, we took measurements in the pulsed mode, where only a short (~few ms) pulse is applied to the gate electrode.[22] No bias stress or long-lived threshold voltage shift were observed in the pulsed mode. Low currents (<100 pA) are not easily measured with the pulsed gate method: the sub-threshold region and the onset voltage of the devices were therefore accurately determined only in the DC mode.

The TFT characteristics are described by an onset voltage when the current rises rapidly from the off-state leakage value, and a threshold voltage, $V_T$, denoting, ideally, that the FET is fully turned on. The onset voltage is typically very close to 0 V indicating little residual doping in the material. A significant voltage between onset and threshold is a characteristic of disordered materials and is due to the localized states. Particularly at low temperature, it is not easy to identify $V_T$, and the choice of $V_T$ affects the apparent mobility. We discuss this issue in more detail below and compare alternative analyses of the data.

2-2 Structural modifications of the poly(thiophene).

The microstructure of the poly(thiophene) material was modified by varying the processing conditions. The TFTs were first characterized immediately after spin-coating and drying the polymer film ("as-spun" devices). The film was then annealed at 80 ºC for one hour followed by 140 ºC for twenty minutes and then cooled to room temperature at approximately 1 ºC/min ("annealed" devices). We also characterized the TFTs after quenching the film from 150 ºC in liquid $N_2$ ("quenched" devices). This temperature was chosen because it corresponds to the isotropic temperature of the polymer, as confirmed by a blue shift in its absorption spectrum and a rapid degradation of its electrical properties. The electrical degradation was reversible upon slow cooling to room temperature.

These different processing conditions altered the crystallinity of the poly(thiophene) film. X-ray spectra of the as-spun and annealed films are shown in Fig. 1. The as-spun film shows a broad diffraction peak at an interplanar spacing of approximately (18.8±0.3) Å. Indeed, solution-processed poly(thiophenes) are known to spontaneously form polycrystalline films upon drying.[2] After annealing, the main diffraction peak narrows and shifts to slightly smaller spacing (17.8±0.3 Å), and a second order diffraction peak is also detectable in this spectrum. The results are consistent with the expectation that the annealed film has a higher degree of structural order. The quenched film did not display any x-ray diffraction peak, and is apparently amorphous.

The results of the room temperature TFT measurements for the three different structural modifications are shown in Table 1, and were reproducible over several samples. The annealed film has the highest mobility followed next by the quenched and then by the as-spun films, with an overall difference of a factor 20-25. The large difference between the onset and threshold in all the films indicates that the Fermi level moves through a substantial population of shallow donor-like states in the polymer as the devices turn on. It is notable that the quenched devices, whose polymer structure lacks crystalline order, have a higher mobility than the as-spun devices. As discussed further below, the transistor action occurs within about 1nm of the dielectric interface and therefore bulk measurements of structural order do not necessarily correlate with TFT performance. For instance, the chemically functionalized dielectric/semiconductor interface may perturb locally the ordering of the polymer.

2-2 Transport measurements as a function of temperature

Low temperatures present both measurement and analysis challenges. At moderate gate voltage, no significant bias stress effects are observed in the DC measurement mode. The onset voltage of the device remains independent of temperature, as observed by Meijer et al.[23] At the lower temperatures, however, a high gate bias is helpful to obtain reliable current measurements. In these conditions, pulsed measurements are used to avoid bias stress and guarantee the stability of the onset voltage. As the temperature is reduced, the on-current – and consequently the mobility – decreases and the turn-on of the device is slower, as shown in Fig. 2a. There is an increase in the sub-threshold slope, and the threshold voltage extracted from the linear regime appears to become more negative with decreasing temperature (see Fig. 2b).

As a first approximation, we extracted the mobility from the linear regime fit of the drain current as a function of gate voltage.[25] The temperature dependence of the mobility of "as spun", "annealed" and "quenched" devices is shown in Fig. 3. In spite of the large differences in room temperature mobility, the general shape of all the curves is the same. The temperature dependence of the mobility of PQT-12 transistors is not monotonic. We interpret the plateau between 220 K and 250 K and the sudden mobility increase at T>250 K as due to structural relaxation in the polymer. A pronounced hysteresis in the current vs. temperature measurements is also observed in this region. A more detailed analysis of this behavior will be published elsewhere. At T<200 K, the temperature dependence of the mobility follows an Arrhenius-like law, with an apparent activation energy, $E_A$, of 40-60 meV and no hysteresis is observed. This activation energy is compatible with literature values for other high-performance polymer semiconductors[2] and indicates that only the temperature dependence of charge transport is measured in this regime.

Inspection of the transfer curves at T<200 K in Figure 4, reveals that a substantial curvature is present, even at high gate voltage, and therefore that the effective mobility of the device depends on gate voltage in addition to temperature. Dimitrakopoulos et al.[26] verified that the gate voltage dependence in organic TFTs is due to the varying charge density and not the gate field. According to standard FET theory, the mobility in the linear regime, as a function of temperature and charge density is:

$$\mu(T,N) = \frac{L \cdot I_{DS}(T)}{W \cdot V_{DS} \cdot N} \quad (1)$$

where L and W are respectively the length and width of the TFT channel and N is the charge density accumulated in the channel. When the TFT is in its on-state, the charge density in the channel is $C_0|V_G - V_T|$ (where $C_0$ is the gate dielectric capacitance, $V_G$ is the gate voltage and $V_T$ is the threshold voltage). The operational definition of the threshold voltage consists of extrapolating the linear portion of the $I_{DS}$ vs. $V_G$ curve to $I_{DS}=0$, in the limit where a small $V_{DS}$ is applied. The intercept at $I_{DS}=0$ is equal to $V_T + \frac{V_{DS}}{2}$. When the mobility depends on charge density, this definition of $V_T$ cannot be applied since $I_{DS}$ does not depend linearly on $V_G$. Physically, however, $V_T$ is the gate voltage at which enough charge is induced in the channel to allow the TFT to be strongly in its on-state. In our measurements, the turn-on voltage of the devices did not shift as a function of temperature. As a consequence, the amount of charge in the channel is determined only by the gate voltage and is independent of temperature. Thus, when the transistor is in its on-state, we have at all temperatures:

$$\mu(T,V_G) = \frac{L \cdot I_{DS}(T)}{W \cdot V_{DS} \cdot C_0(V_G - V_T^0)} \quad (2)$$

where $V_T^0$ is the threshold voltage measured at room temperature. Eq. 2 is valid when the device is switched on strongly (i.e. $V_G \gg V_T^0$) and allows to extract the effective mobility from transfer measurements without having to calculate derivatives of $I_{DS}(V_G)$.

An example of µ as a function of 1/T at different $V_G$, using eq. 2, is shown in Fig. 5a for the annealed devices. The dependence is Arrhenius-like with the apparent activation energy decreasing with increasing $V_G$. The fitted lines intercept the y-axis approximately at the same point (µ~0.6 cm$^2$/V s). We do not observe the intersection at a finite temperature (Meyer-Neldel rule) as reported by Meijer et al. in similar measurements of other organic semiconductors (P3HT, PTV, and pentacene).[27,28] Figure 5b is similar to Fig. 5a comparing the mobility data measured in the three types of films. For clarity, not all the experimental data points are shown here. Interestingly, the extrapolated mobility at 1/T=0 for the three films varies only by a small factor even though the room temperature mobility spans almost two orders of magnitude.

Comparison of the mobility values in Figs. 3 and 5 show a similar magnitude and temperature dependence. Since the two different methods of analysis give similar results, we conclude that the uncertainty in analysis does not affect the broad features of the result.

**3-Charge transport and device models**

In this section we fit the temperature dependence data to models of the electronic structure and transport properties of the polymer. The variable range hopping model proposed by Vissenberg and Matters[29] is compared with a multiple ME model,[30] and we argue that the ME model is preferred. The parameters extracted from this model are then used in a device simulation program in order to compare the calculated transfer curves to the experimental ones. Although the ME model is simplified and should be viewed as semi-quantitative, we show that it captures sufficiently well the transport characteristics of the polymer and provides an insight in the effect of processing and microstructure.

3-1 The Vissenberg and Matters hopping model[29]

Vissenberg and Matters developed a transport model based on variable range hopping in an exponential DOS and including percolation. This model was developed to analyze the temperature dependence of charge transport in organic semiconductors, and was primarily intended to apply to amorphous polymers. The authors describe the conductivity in the polymer as equivalent to transport through a resistor network where the nodes of the network have different energies according to the exponential DOS. The percolation criterion through the network is then related to the temperature, the position of the Fermi level and the width of the exponential tail of the DOS. In this model, charge transport occurs by variable-range hopping. The calculated field-effect mobility is,

$$\mu = \frac{\sigma_0}{e}\left[\frac{\pi\left(\frac{T_0}{T}\right)^3}{(2\alpha)^3 B_c \Gamma\left(1-\frac{T}{T_0}\right)\Gamma\left(1+\frac{T}{T_0}\right)}\right]^{\frac{T_0}{T}} \times \left[\frac{(C_0 V_G)^2}{2kT_0\varepsilon_s}\right]^{\frac{T_0}{T}-1} \qquad (3)$$

where $\sigma_0$ is the conductivity prefactor, $\alpha$ is the wavefunction overlap parameter, $kT_0$ is the width of the exponential tail of the DOS, $\varepsilon_s$ is the dielectric constant of the semiconductor, $\Gamma(x) = \int_0^\infty \exp(-y) y^{x-1} dy$ ) and $B_c$ is a constant (~2.8). The fitting parameters are $\sigma_0$, $\alpha$ and $T_0$. In this model, the mobility is essentially governed by the width of the DOS and the overlap parameter. Formally, the conductivity pre-factor represents the limit of the conductivity as 1/T tends to 0. This model however does not apply when 1/T tends to 0 therefore it is difficult to assign a physical meaning to the conductivity pre-factor. Vissenberg and Matters modeled the temperature-dependence of the field-effect mobility of polythienylene vinylene (PTV) and solution-processed pentacene. The fitted DOS tail width that they obtained from their devices was approximately 33 meV ($T_0$=380 K) for both pentacene and PTV, even though the room temperature mobility of these materials differs by over two orders of magnitude. The authors attribute the mobility difference to the different tunneling rate between sites (i.e. to different values of the overlap parameter).

Eq. 3 was used to fit our mobility vs. T data for the three structural modifications of PQT-12, with $\sigma_0$, $\alpha$ and $T_0$ as free parameters. The best fit to the data is obtained by allowing $\sigma_0$ to vary over a wide range. Since the physical meaning of this parameter is questionable, we prefer to fit our mobility vs. T data by varying $\sigma_0$ as little as possible. Acceptable fits are obtained with the parameters listed in Table 2 along with those obtained by Vissenberg and Matters for PTV and pentacene. According to this model, the transport differences in our PQT-12 films are due mostly to differences in the overlap parameter $\alpha$. Since the overlap parameter is directly related to the wavefunction decay, it is not obvious why it would vary in microstructural modifications of the same polymer. We thus also attempted to fit our data with this model by keeping $\alpha$ constant ($\alpha^{-1}$=1Å ). The fit is visibly worse than in the previous cases. No correlation can be drawn between $T_0$, which should be an indication of disorder in the material, and the device mobility (Table 2). Moreover, the conductivity pre-factor varies markedly between the three devices.

We therefore conclude that it is doubtful that our results can be explained by the Vissenberg and Matters hopping model and that a physical meaning can be attributed to the fitted parameters.

3-1 Mobility edge (ME) model

The ME model assumes that there is a defined energy (the mobility edge) in the DOS that separates mobile states from localized states. Trapped carriers become temporarily mobile by thermal excitation to the mobile states. The ME model applies intuitively to polycrystalline materials, where the mobile states are extended band states of the crystallites and the trapped states are located in the disordered regions between the crystallites. Hopping directly between trapped states is a competing transport mechanism.[31] Given our material properties however, at time scales longer than approximately 1 µs, transport through thermal excitation to mobile states is expected to dominate.[31] Thus, our experimental conditions are such that we can neglect direct hopping between trapped states.

In order to apply the ME model, we assume that the DOS of the polymer is described by bands with tails extending in the gap of the form shown in Fig. 6. The essential property of this DOS is that it varies "slowly" with energy inside the band near the band edge ($D(E) \sim E^n$)[10] while it varies exponentially with energy inside the band gap. We define the mobility edge, E=0, at the top of the band-like states.

The assumption of an exponential tail is justified as an approximation of the Gaussian distribution generally accepted for organic semiconductors. Moreover, completely random disorder is not likely to occur in a polycrystalline material such as PQT-12, which shows texture by X-ray diffractometry, making a fully Gaussian DOS less preferable than in amorphous polymers. Finally, a DOS with an exponential tail has been successfully used to model temperature dependence of conductivity in other disordered materials such as hydrogenated amorphous silicon.[32-35] Thus, the localized donor-like states are represented by an exponential DOS that decays into the band gap as,

$$D_{exp} = \frac{N_{tot}}{E_b} e^{-\frac{E}{E_b}} \tag{4}$$

where $N_{tot}$ is the total concentration of tail states and $E_b$ is the width (in eV) of the exponential tail. Holes in this part of the DOS are trapped and do not contribute to the transistor current.

The exact functional form of the band DOS is not known in polymer semiconductors. Here we use the simplest form for a free electron gas in three dimensions, where the DOS is proportional to $E^{1/2}$. Holes at energies below the mobility edge $E_0$ are mobile and assumed to have a constant mobility $\mu_0$, while holes located at energies above the mobility edge have zero mobility.

As the gate bias varies, the Fermi level of the semiconductor moves in the DOS as a result of charge accumulated in the polymer. Taking the simple approximation of a fixed depth of the conducting channel, the Fermi energy $E_F$ in the polymer semiconductor satisfies the following equation:

$$N_{tot}(V_G) = \frac{C_0 |V_G - V_T|}{h} = \int_{-\infty}^{+\infty} D(E) f(E_F, E) dE \qquad (5)$$

where $h$ is the channel dimension normal to the dielectric surface (assumed to be 1 nm), $f(E_F,E)$ is the Fermi-Dirac distribution, and $D(E)$ is the DOS of the polymer. Solving Eq.5 numerically allows us to determine $E_F(V_G,T)$. The concentration of mobile carriers is then:

$$N_{mob}(V_G, T) = \int_{-\infty}^{E_0 \equiv 0} D(E) f(E_F, E) dE \qquad (6)$$

and the effective mobility is:

$$\mu(V_G, T) = \mu_0 \frac{N_{mob}}{N_{tot}} \qquad (7)$$

3-2 ME model fit to the data

A density of states distribution $D(E)$ based on eqs 5-7 that fits the whole set of temperature dependent data is found by successive iterations using $N_{tot}$, $E_b$ and $\mu_0$ as fitting parameters. The parameter values are shown in Table 3 and the fit to the temperature dependence of the mobility is shown in Fig. 7a-c, for the three different structural modifications. A good fit to the data is obtained, and parameters show that the band tail width is the main difference between the structural modifications.

The above analysis assumes uniform conduction in a fixed channel. In a TFT, however, the effective mobility is a function of the charge density, which decreases as a function of distance from the semiconductor/dielectric interface. In order to improve the model, and to simulate our I-V data, a transistor model was used.[36] Briefly, this is a 1-dimensional model where the current in a TFT is calculated within the gradual channel approximation, in the linear regime. For each value of the gate voltage, the charge and potential distributions are calculated in the polymer as a function of distance from the semiconductor/dielectric interface. Because this is a 1-D model, it could not be used to simulate the full behavior of the device and in particular the saturation regime. Also, we chose not to model the sub-threshold behavior of the device as it depends on subtle details of the DOS. The purpose of these simulations was to validate the results obtained by applying the ME model rather than to provide a full model of the device behavior. The input of the model is the DOS of the polymer and the calculation is compared to the I-V characteristics of the devices at three temperatures: 180 K, 140 K, 90 K, and the results are shown in Figure 8a-c. In each case, once the material parameters are fixed by the fit at 180 K, good agreement between simulation and experiment is also obtained at the other temperatures. The parameters used for the simulation are shown in Table 3. The band mobility and the band tail widths are similar to those obtained by the simple ME model. The difference in the total concentration of tail states, which is less than a factor of 3, reflects the different assumptions of the two models. We conclude that the parameters are not very sensitive to the details of the model assumptions.

**4- Discussion**

4-1 Transport models

The ME model is widely used for inorganic amorphous and polycrystalline materials.[32,33,37,38] Horowitz et al. also use it to explain the temperature and gate voltage dependence of polycrystalline substituted and unsubstituted oligothiophene films.[39-41] These authors approximate the DOS of their films with a double exponential. The shallow tail state concentration was $2 \times 10^{20}$ cm$^{-3}$ and the width was 10.3 meV (120K) for both an unsubstituted and an end-substituted sexithiophene. Their estimates of the intrinsic mobility of the unsubstituted sexithiophene vary between 0.03 cm$^2$/V s and 50

cm$^2$/V s. The ME model is also successful in interpreting pentacene TFT measurements with a similar DOS.[36]

While it is easy to justify the existence of a mobility edge separating extended mobile states from localized band tails in amorphous covalently bonded semiconductors (e.g. amorphous silicon) or in polycrystalline inorganic and organic semiconductors, this assumption may be more controversial in polymer films. In conjugated polymers, it is usually assumed that charge transport is limited by π-π intrachain coupling. Charges are delocalized within a conjugation length along the chain but are laterally confined in the polymer chain. A hopping model between chains seems the appropriate mechanism to account for bulk transport. Furthermore, polymers exhibit polaron behavior, in which holes are self-localized by structural relaxation, and the polaron binding energy can be enhanced at distortions in the polymer structure. Polarons are associated with low mobility transport, and with a hopping mechanism having moderately high activation energy.

However, the high mobility regio-regular poly(thiophenes) such as PQT-12 or poly(3-hexylthiophene) (P3HT) do form polycrystalline films with strong π-π coupling within the crystallites. Our X-ray diffraction spectra show that the PQT-12 films certainly have a polycrystalline microstructure. Brown et al.[42] and Österbacka et al.[43] found clear evidence of interchain delocalization of the charge carriers in polycrystalline poly(thiophene) films, thus indicating the existence of an extended mobile state within the crystallites. These studies also show that the polarons that are evident in the regio-random amorphous forms of poly(thiophene), are absent in the regio-regular polycrystalline forms. We therefore consider that a transport mechanism involving a mobility edge and a distribution of shallow traps is a reasonable model for the high mobility materials even if a hopping model applies to lower mobility amorphous polymers.

The ME and hopping models are actually difficult to distinguish purely based on the fit to the experimental data. The mobility calculated with the hopping model of Eq. 3 has an Arrhenius-like temperature dependence within the restricted temperature range over which measurements of organic TFTs are typically made. Our preference for the ME model is based on the expected transport mechanism of polycrystalline films, and on the unphysical parameter extraction for the hopping model. Since both models share the same functional dependence of mobility on temperature, we applied our model to fit the PTV data used by Vissenberg and Matters and the results are shown in Table 3.[44] Similar values as PQT-12 are obtained for the intrinsic mobility and the total concentration of donor states. The lower mobility of PTV ($\sim 2 \times 10^{-3}$ cm$^2$/V s) compared to PQT-12 is primarily attributed to the width of the exponential tail of the DOS. We therefore suggest that the ME model may be more appropriate for PTV.

4-2 Parameter extraction

The ME model discussed here is simplified but nevertheless captures the essential features of the temperature and gate voltage dependence of the transport properties of PQT-12. The key assumptions are that a reasonably sharp mobility edge can be defined in this material[10] and that the DOS can be modeled as a slowly varying distribution of band states followed by a rapidly varying tail. The values of $N_{tot}$, $E_b$, and $\mu_0$ are bound by physical requirements. The free carrier mobility, $\mu_0$, extracted from the model, is larger than the intersection point of the Arrhenius-like fits of the data at 1/T=0 ($\sim 0.6$ cm$^2$/V s) because the statistical shift (i.e. $\frac{\partial E_F}{\partial T}$) must be negative in this case, where $E_F$ moves towards the lower DOS as the temperature is increased.[30] In order to correctly fit the experimental data, $E_b$ must be of the order of $E_A$, the activation energy in the Arrhenius-like fit of the experimental results. Finally, $N_{tot}$ must be such that at the charge concentrations imposed by the experimental gate voltages, the Fermi level stays in the exponential tail of the distribution in order to insure Arrhenius-like temperature dependence of the effective mobility. Although we are obviously not able to explore the whole parameter space and cannot guarantee that the parameters shown in Table 3 are a unique set, their values satisfy the previously listed requirements and are physically reasonable. Finally, if intrinsic mobilities higher than $\sim 4$ cm$^2$/V s are used to fit the data (using $D(E) \sim E^{1/2}$ for the band DOS), unrealistic DOS shapes are obtained (i.e. the DOS in the tail is higher than in the band).

The position of the mobility edge inside the valence band and the exact shape of the valence band edge are not known, and were not varied in the numerical model. The position of the Fermi level (and therefore the fraction of trapped charges) depends logarithmically on the DOS at the band edge and is therefore not extremely sensitive to its absolute value. We verified that the parameters extracted from the simulations remain within the same order of magnitude and show the same qualitative trends for n=0 (2-D hole gas), ½ (3-D hole gas) and 1 (the approximate DOS exponent in hydrogenated amorphous silicon).[30]

The ME model indicates that the free carrier mobility is ~1 $cm^2/V$ s, and the uncertainties of our model allow us to give only an estimate of ~1-4 $cm^2/V$ s for the upper limit of the mobility of the PQT-12 films. It is interesting to consider if the measured value corresponds to the mobility in a "single crystal" transistor made with PQT-12, or whether grain size scattering effects limit the free carrier mobility in the polycrystalline films. To date, the highest mobility measured in an organic single crystal is 8 $cm^2/V$ s.[45]

The intrinsic carrier mobility is the upper bound of the field-effect mobility of all PQT-12 films whose ordered regions share the same structure as the ones examined here. In order to obtain the highest mobility polymer TFT however, it is not sufficient to design the molecular structure with the objective to maximize the intrinsic mobility. The microstructure of the film must also be such as to reduce the band tail energetic disorder as much as possible in order to minimize the number of trapped carriers.

4-3 Structural modifications

The room temperature mobility of the devices examined here varied by a factor of 20-25, and even more at low temperature. The ME model indicates that the difference originates mostly from the different widths of the exponential tail of the DOS. In other words, for our material and within our range of processing conditions, the mobility is governed essentially by the amount of trapping in shallow donor-like tail states. Indeed, the intrinsic mobility is approximately the same for all the devices examined here. A wider band tail is associated with a more disordered material that will therefore have a lower mobility. This result agrees with the view that at the dielectric interface the films are all made of ordered regions having approximately the same structure, which governs the intrinsic mobility, and disordered regions, which govern the trap states. The three films processed differently have different distributions of trap states, which lead to different mobility. Nevertheless, the films must have essentially the same structure in the ordered regions since they have the same intrinsic mobility. It is not surprising that the as-spun film has the broadest DOS tail, since we would expect to find the largest variations in local structure in a film whose structure has not been allowed to equilibrate through an annealing step, in agreement with the x-ray diffraction data.

It was noted earlier that the X-ray diffraction of the quenched film indicates an amorphous structure, even though the mobility is higher than the as-spun film. The parameters extracted from the ME model suggests that the interface region where transport occurs actually has an ordered polycrystalline structure. It seems to us reasonable to suppose that the polymer material immediately adjacent to the dielectric has an ordered structure largely determined by the interface, even if the bulk material is disordered by the high temperature anneal and subsequent rapid quenching.

## 5-Conclusions

The field-effect mobility of polymer TFTs made with PQT-12 as the active material shows a thermally activated behavior below 200 K. The apparent activation energy decreases with increasing charge concentration in the channel. These results can be well understood in terms of trapping of carriers in localized states in the exponential tail of the DOS and thermal excitation in mobile states. The room temperature field-effect mobility of the TFTs can vary by a factor of ~20-25 as a result of different processing conditions. According to the ME model, however, these differences are mostly due to differences in the width of the exponential tail of the DOS, and not in the intrinsic mobility in the ordered regions of the material: the lower the mobility, the wider the donor tail width. The width of the exponential tail of the DOS is usually associated with the degree of structural order of the material. We estimate the intrinsic mobility of our PQT-12 films to be ~1-4 $cm^2/V\,s$.

Even though our simple model captures the essential features of the PQT-12 transistors, more accuracy in modeling the charge transport properties of the devices would be obtained by introducing a more realistic DOS. In this regard, experimental determination of the DOS of polymer thin films is of key importance.

**Acknowledgements.**

The authors gratefully acknowledge R. B. Apte, W. S. Wong, M. L. Chabinyc, J. P. Lu and J. E. Northrup of PARC for helpful discussions and B. Krusor for help with the X-ray spectra. This work is partially supported by the Advanced Technology Program of the National Institute of Standards and Technology (contract 70NANB0H3033)

**List of figures**

**List of Tables**

**Table I:** Typical room temperature electrical characteristics of the PQT-12 films. Occasionally devices stored for a long time in an inert atmosphere showed a smaller sub-threshold slope (~0.5 V/dec) and consequently a lower $V_T$(~-5V) as the one shown in Fig. 2.

**Table II:** Parameters extracted from fitting the PQT-12 data to the Vissenberg and Matters hopping model.

**Table III:** Parameters extracted from fitting the PQT-12 data to the ME model (left panel). Parameters used for TFT simulation (right panel).

|          | Mobility (cm$^2$/V s) | $V_{on}$ (V) | $V_T$ (V) | Sub-threshold slope (V/dec) |
|----------|-----------------------|--------------|-----------|------------------------------|
| Annealed | $0.09 \pm 0.01$       | $0 \pm 1$    | $-9 \pm 2$ | $0.8 \pm 0.2$ |
| Quenched | $0.02 \pm 0.01$       | $0 \pm 1$    | $-12 \pm 1$ | $0.9 \pm 0.2$ |
| As spun  | $(4 \pm 1) \times 10^{-3}$ | $0 \pm 1$ | $-12 \pm 1$ | $2.7 \pm 0.4$ |

**Table I:** Typical room temperature electrical characteristics of the PQT-12 films. Occasionally devices stored for a long time in an inert atmosphere showed a smaller sub-threshold slope (~0.5 V/dec) and consequently a lower $V_T$(~-5V) as the one shown in Fig. 2.

|  | 3 free parameters fit | | | 2 free parameters fit | | |
|---|---|---|---|---|---|---|
|  | $\sigma_0$ (S/m) | $kT_0$ (meV) | $\alpha^{-1}$ (Å) | $\sigma_0$ (S/m) | $kT_0$ (meV) | $\alpha^{-1}$ (Å) |
| Annealed | $3.5 \times 10^{11}$ | 27.6 | 1.28 | $1.9 \times 10^{12}$ | 28.0 | 1 |
| Quenched | $3.9 \times 10^{11}$ | 29.3 | 1.05 | $5.8 \times 10^{11}$ | 29.3 | 1 |
| As spun | $3.7 \times 10^{11}$ | 31.9 | 0.89 | $7.2 \times 10^{10}$ | 28.4 | 1 |
| PTV (from Ref 29) | $7 \times 10^{9}$ | 33 | 0.8 | - | - | - |
| Pentacene (from Ref 29) | $1.6 \times 10^{10}$ | 33 | 2.2 | - | - | - |

**Table II:** Parameters extracted from fitting the PQT-12 data to the Vissenberg and Matters hopping model.

|  | Parameters extracted from ME model | | | Parameters used in device simulation | | |
|---|---|---|---|---|---|---|
|  | $\mu_0$ (cm$^2$/V s) | $E_b$ (meV) | $N_{tot}$ (cm$^{-3}$) | $\mu_0$ (cm$^2$/V s) | $E_b$ (meV) | $N_{tot}$ (cm$^{-3}$) |
| Annealed | 0.8 | 34 | 6.5x10$^{20}$ | 0.75 | 33 | 1.5x10$^{21}$ |
| Quenched | 0.8 | 40 | 8x10$^{20}$ | 0.75 | 45 | 1.5x10$^{21}$ |
| As spun | 1 | 50 | 5.5x10$^{20}$ | 0.75 | 58 | 1.5x10$^{21}$ |
| PTV (from Ref 29) | 0.6 | 45 | 7.2x10$^{20}$ | - | - | - |

**Table III:** Parameters extracted from fitting the PQT-12 data to the ME model (left panel). Parameters used for TFT simulation (right panel).

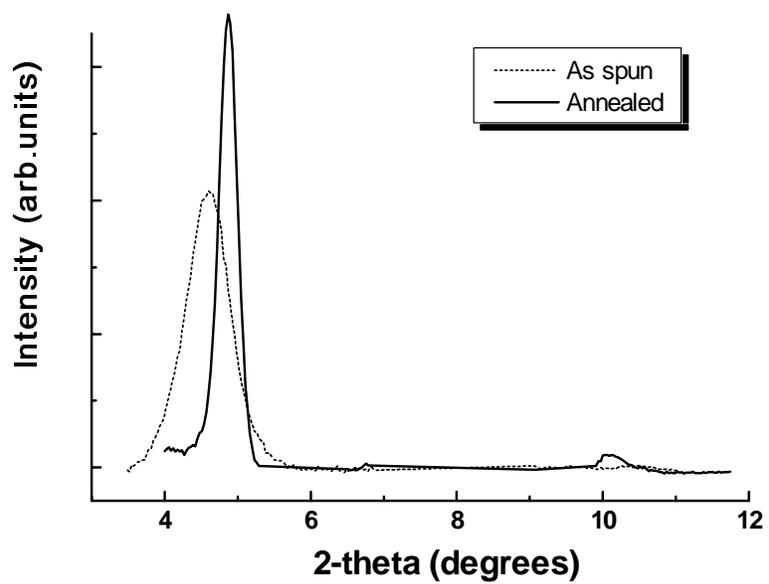

Figure 1: Salleo et al. "Intrinsic hole mobility and trapping in a regio-regular poly(thiophene)

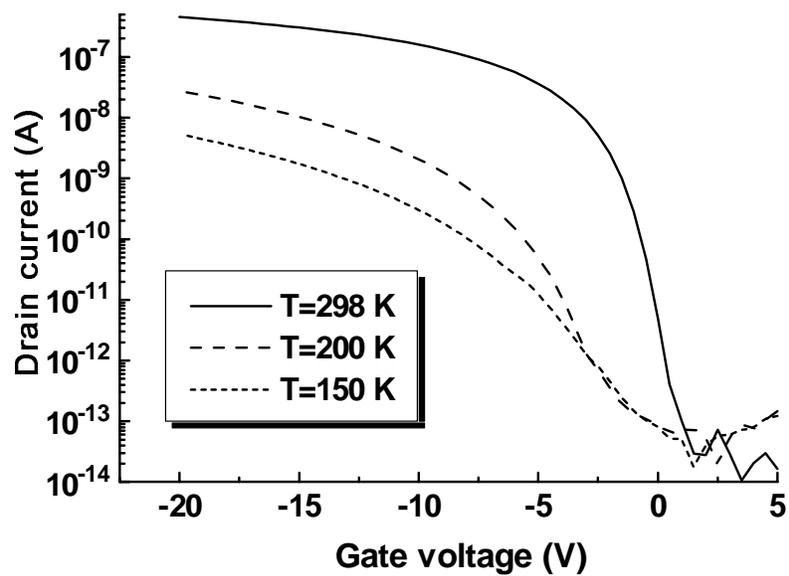

Figure 2a: Salleo et al. "Intrinsic hole mobility and trapping in a regio-regular poly(thiophene).

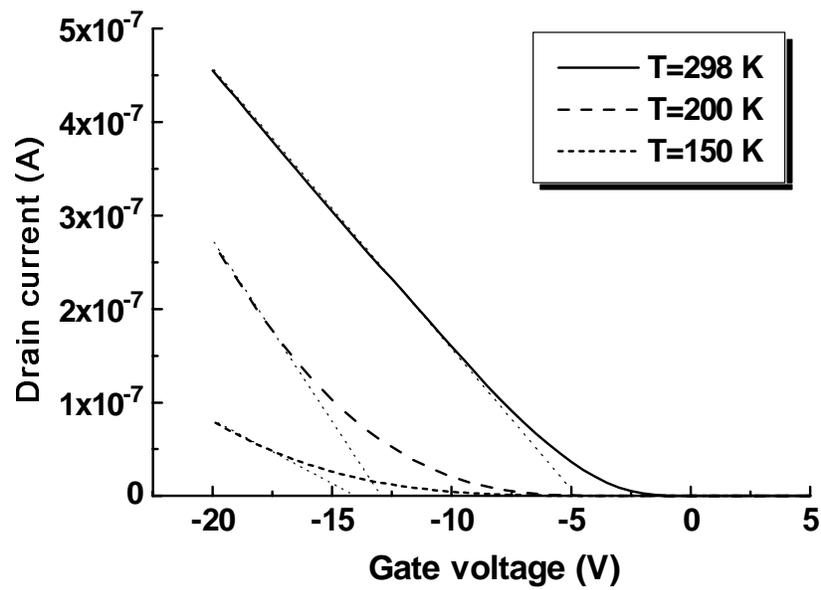

Figure 2b: Salleo et al. "Intrinsic hole mobility and trapping in a regio-regular poly(thiophene).

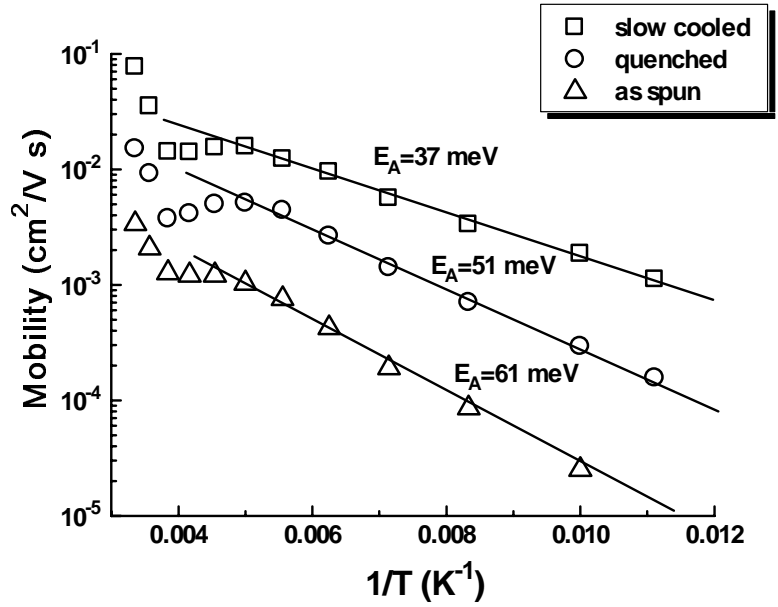

Figure 3: Salleo et al. "Intrinsic hole mobility and trapping in a regio-regular poly(thiophene).

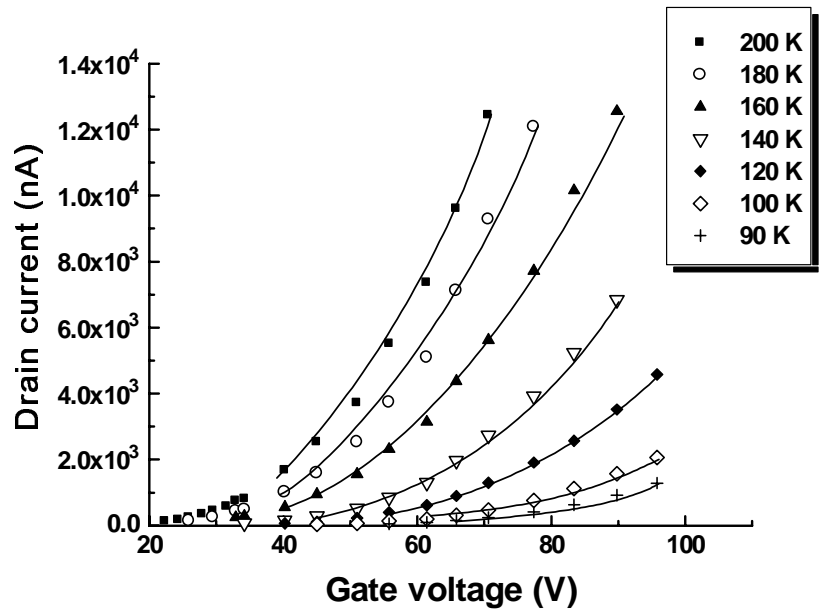

Figure 4: Salleo et al. "Intrinsic hole mobility and trapping in a regio-regular poly(thiophene).

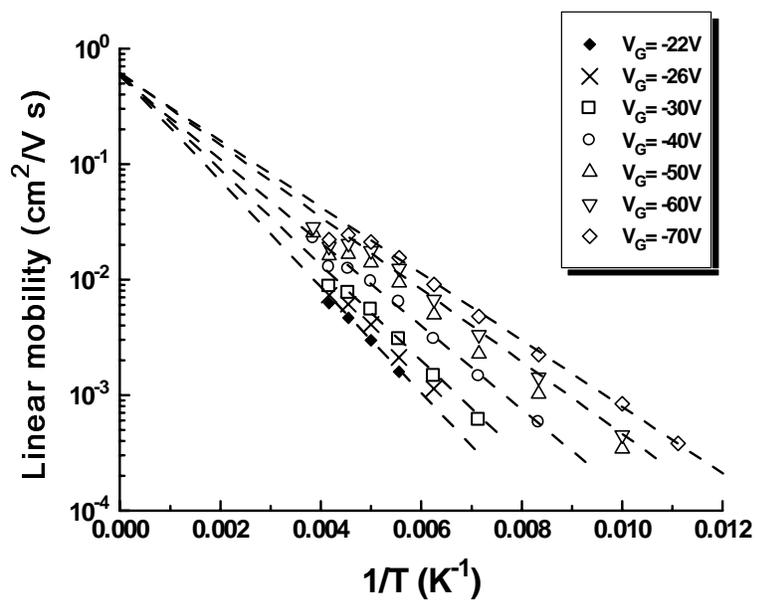

Figure 5a: Salleo et al. "Intrinsic hole mobility and trapping in a regio-regular poly(thiophene).

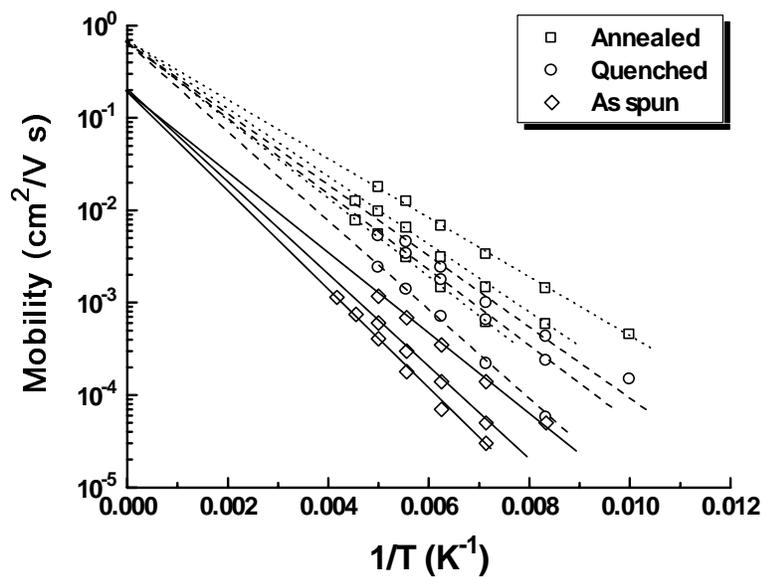

Figure 5b: Salleo et al. "Intrinsic hole mobility and trapping in a regio-regular poly(thiophene).

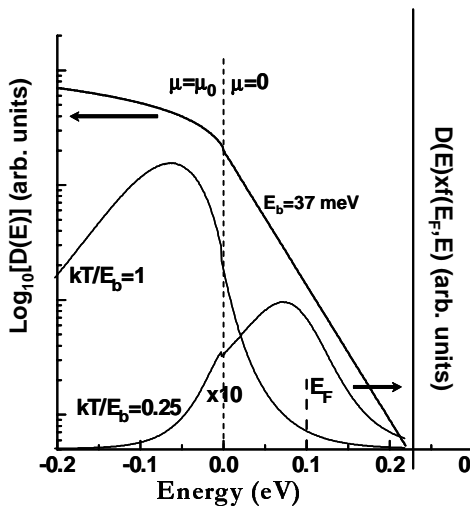

Figure 6: Salleo et al. "Intrinsic hole mobility and trapping in a regio-regular poly(thiophene).

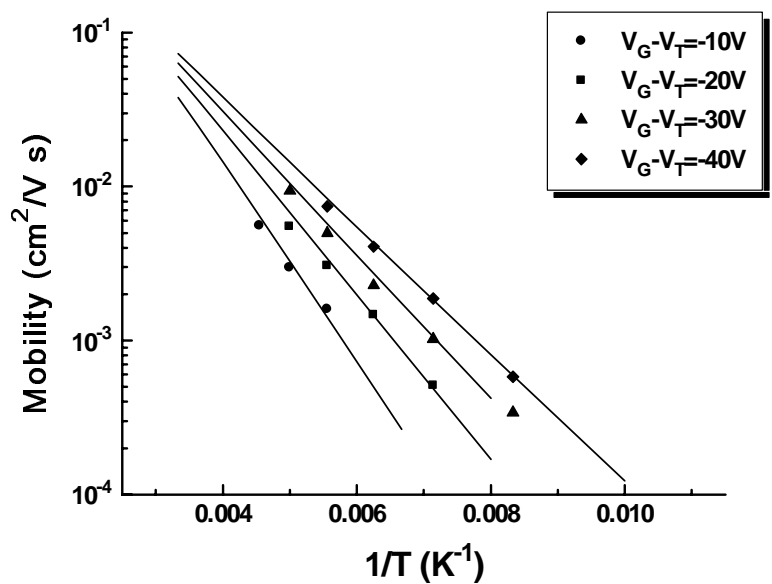

Figure 7a: Salleo et al. "Intrinsic hole mobility and trapping in a regio-regular poly(thiophene).

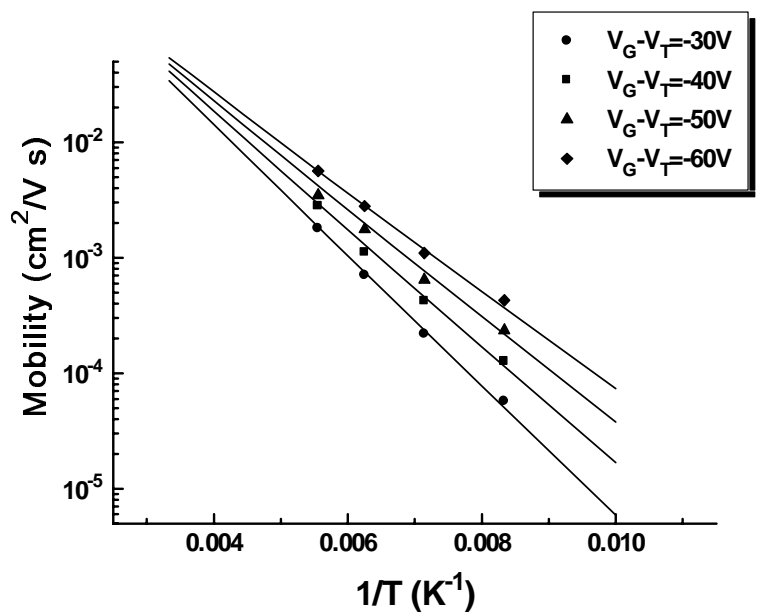

Figure 7b: Salleo et al. "Intrinsic hole mobility and trapping in a regio-regular poly(thiophene).

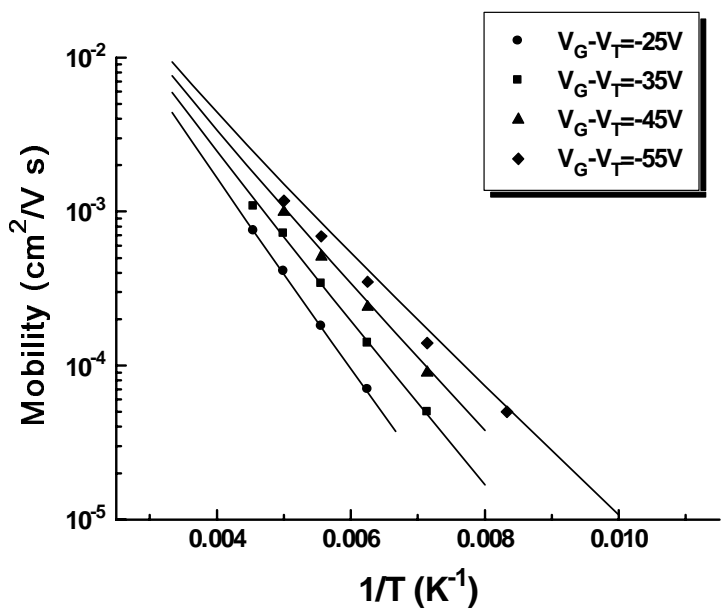

Figure 7c: Salleo et al. "Intrinsic hole mobility and trapping in a regio-regular poly(thiophene).

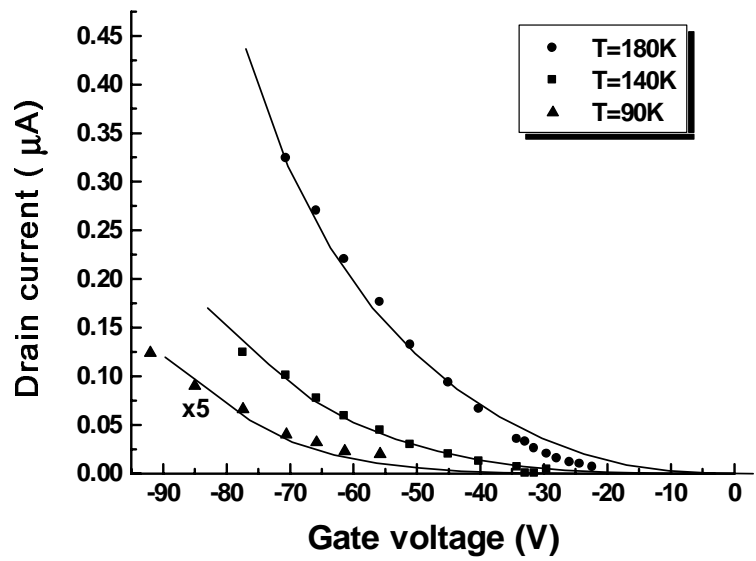

Figure 8a: Salleo et al. "Intrinsic hole mobility and trapping in a regio-regular poly(thiophene).

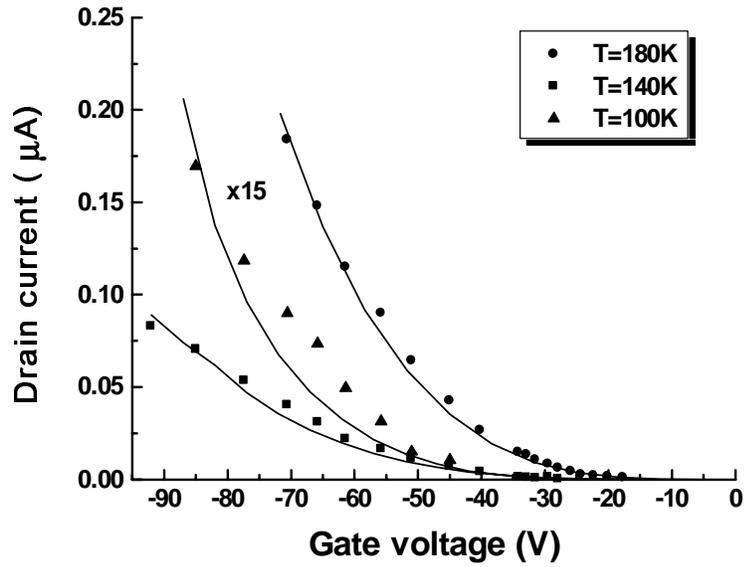

Figure 8b: Salleo et al. "Intrinsic hole mobility and trapping in a regio-regular poly(thiophene).

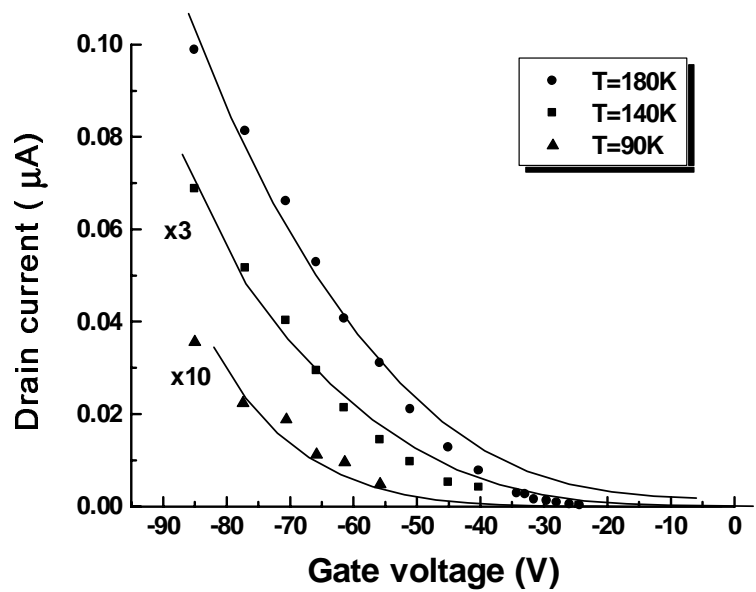

Figure 8c: Salleo et al. "Intrinsic hole mobility and trapping in a regio-regular poly(thiophene).